\documentclass[10pt,conference]{IEEEtran}
\IEEEoverridecommandlockouts
\usepackage{cite}
\usepackage{amsmath,amssymb,amsfonts}
\usepackage{algorithmic}
\usepackage{graphicx}
\usepackage{textcomp}
\usepackage{color}
\def\BibTeX{{\rm B\kern-.05em{\sc i\kern-.025em b}\kern-.08em
    T\kern-.1667em\lower.7ex\hbox{E}\kern-.125emX}}
    
\renewcommand{\baselinestretch}{0.89}
\begin{document}

\title{A Reputation-based Stackelberg Game Model to Enhance Secrecy Rate in Spectrum Leasing to Selfish IoT Devices\\
}
%


\author{
   \IEEEauthorblockN{
    Fatemeh Afghah\IEEEauthorrefmark{1},
     Alireza Shamsoshoara\IEEEauthorrefmark{1},
     Laurent Njilla\IEEEauthorrefmark{2}, 
    and
     Charles Kamhoua\IEEEauthorrefmark{3}}
   \IEEEauthorblockA{
     \IEEEauthorrefmark{1}School of Informatics, Computing and Cyber Systems,
    Northern Arizona University, Flagstaff, AZ, United States, \\Email: \{fatemeh.afghah,alireza\_shamsoshoara\}@nau.edu}
   \IEEEauthorblockA{
    \IEEEauthorrefmark{2}Air Force Research Laboratory,
    Rome, NY, United States, Email: laurent.njilla@us.af.mil}
 \IEEEauthorblockA{
    \IEEEauthorrefmark{3}
    Army Research Laboratoy, Adelphi, MD, Email: charles.a.kamhoua.civ@mail.mil}
 }

\maketitle
\begin{abstract}
The problem of cooperative spectrum leasing to unlicensed Internet of Things (IoT) devices is studied to account for potential selfish behavior of these devices.
A distributed game theoretic framework for spectrum leasing is proposed where the licensed users can willingly lease a portion of their spectrum access to unlicensed IoT devices, and in return the IoT devices provide cooperative services, firstly to enhance information secrecy of licensed users via adding intentional jamming to protect them from potential eavesdroppers, and secondly to enhance the quality of communication through cooperative relaying. The cooperative behavior of the potentially selfish IoT devices is monitored using a reputation-based mechanism to enable the primary users to only interact with the reliable IoT devices. 
The simulation results show that using the proposed reputation-based method enhances the secrecy rate of the primary users by reducing the possibility of attacks from selfish IoT
devices. Hence, this model can offer a practical solution for spectrum leasing with mobile IoT devices when assuring the required quality of communication and information secrecy for the spectrum owners\footnote{DISTRIBUTION A. Approved for public release: distribution unlimited. Case
Number: 88ABW-2018-0002. Dated 02 Jan 2018.
}.
\end{abstract}


\section{Introduction}
Development of new wireless communication technologies and skyrocketing demand for new data services, in particular growing Internet of Things (IoT) with potentially billions of wireless devices, motivated many dynamic spectrum allocation technologies to enhance the spectrum utilization efficiency. While current research efforts focus on common spectrum sharing mechanisms where the licensed users are unaware of unlicensed users' presence (e.g. database control and spectrum sensing mechanisms), these methods may not be the best choice for spectrum sharing with IoT devices. Energy constraint and limited spectrum mobility at these often tiny devices restrict their performance in spectrum sensing mechanisms. Moreover, having a central controller or enabler to keep a record of licensed users' spectrum usage and allowable interference range as required in database control technologies (e.g. TV White Spaces) may not be practical in such large-scale IoT networks. 

Another drawback of the majority of common spectrum sharing models is their overly conservative approach to protect the incumbent users against malicious and selfish unlicensed entities. Hence, such approaches that enforce a low level of transmission power for unlicensed users, or allocate a wide and static protection zone around the incumbents, are unable to address the increasing demand for radio spectrum from IoT networks.
Furthermore, in common model paradigm, the licensed users do not usually benefit from allowing the unlicensed users to share their spectrum, but even they can suffer due to interference from Secondary Users (SUs). That being said, flexible spectrum right models in which the Primary Users (PUs) can willingly lease a portion of their spectrum access to unlicensed users (e.g. IoT devices) in exchange for remuneration or some sort of compensation (e.g. energy harvesting or cooperative relaying service) can offer a win-win solution for both parties \cite{Zhang_Leasing_IoT,Simeone, Korenda_CISS,Namvar_CISS,Afghah_CDC,Afghah_NWRCS}.

However, one main concern toward enabling shared access techniques to the radio spectrum, particularly the federal bands, is security. In addition to traditional security threats in wireless networks such as concerns related to privacy of the users, authentication, and confidentiality of communications among the users in the presence of eavesdroppers, networks with dynamic spectrum allocation mechanism are also vulnerable to unauthorized access to the spectrum as well as potential selfish behavior of users taking advantage of the ad hoc nature of such networks \cite{Security_Park,Survey_Attar,razi2017optimal}. Therefore, the cognitive radio networks are prone to suffer from various exogenous, malicious, and selfish attacks. Some of these attacks are Denial of Service (DoS) attacks to deny the unlicensed users from spectrum access, sensing data falsification attacks, and primary user emulation attacks, only to name a few \cite{Survey_Fragkiadakis,Yu_False_sensing}. Different security mechanisms have been proposed to address these attacks, which among these information theoretic secrecy methods aim to secure the communication of the nodes from potential eavesdroppers by exploiting the physical characteristics of wireless channels \cite{Poor_Secrecy}. This can be achieved by adding artificial noise, using beamforming techniques, or employing MIMO techniques \cite{Nguyen_Beamforming,Stanojev_Secrecy}. Despite conventional cryptographic methods, information theoretic secrecy-based techniques do not rely on encryption keys, and hence do not involve key distribution and key management complexities.  

Cooperative jamming in cognitive radio networks has been implemented by employing the SUs to create artificial noise \cite{Liu_Secrecy_Game}, or transmit structured codewords to reduce the eavesdropper's capability in decoding the PU's information. The potentially non-cooperative SUs can be encouraged to provide such a service if they are granted with a chance of spectrum access for their own transmission, as introduced in \cite{Liu_Secrecy_Game}. Providing cooperative jamming for the primary users, in addition to other previously studied compensation techniques such as cooperative relaying and energy harvesting, can offer a practical solution for implementation of property-right cognitive radio networks without involving money exchange among the parties or other regulatory issues \cite{Liu_Secrecy_Game}. In \cite{Stanojev_Secrecy}, a Stackelberg game model is defined to model the interactions between a legitimate source node and a non-altruistic secondary user in presence of an eavesdropper. The secondary user can be compensated with a fraction of the legitimate source node's access time to radio spectrum if it provides cooperative jamming to enhance the secrecy rate of the source node. This model is further extended to a scenario with multiple potential cooperative jammers where the competition among them is modeled as an inner Vickrey auction. The proposed model \cite{Stanojev_Secrecy} favors the primary user by defining the game in such a way that the time allocation to access the spectrum as well as the ratio of the power, which the secondary users spend on cooperative jamming and the one they utilize for their own transmission, are determined by the primary user. Although the primary user deserves more benefits as the spectrum owner, in reality there is no guarantee that the secondary users will follow the power allocation ratio requested by the primary user after being granted with spectrum access. The authors in \cite{Talabani} study a similar model with the objective of enhancing the secrecy rate of both the primary and secondary nodes, where the secondary nodes are assumed to be fully trustable.


While cooperative jamming provided by the SUs can potentially enhance the secrecy rate of the licensed users, it involves the assumption of having unconditionally cooperative and trustable SUs, as considered in the majority of previously reported studies \cite{Liu_Secrecy_Game,Stanojev_Secrecy,Talabani,kamhoua2012surviving}. However, this assumption is far too optimistic in communication networks, noting the non-altruistic nature of cognitive secondary users (e.g. IoT devices). This frequent assumption has been revisited in some work considering different aspects of potential selfish or malicious behavior by SUs. The authors in \cite{He_Yener,Yuksel} discussed a scenario where the relay nodes can potentially act as eavesdroppers and study whether cooperation can still improve the secrecy. In this paper, we consider a scenario where the secondary users can be selfish but not malicious and propose a reputation-based mechanism to identify the untrustable users. 
It is assumed that the SUs are pre-authenticated and the problem of dealing with authenticating intruding/malicious secondary nodes is out of the scope of this work.  

In the majority of previously proposed reputation-based methods to enforce cooperation, the users (e.g. relay nodes) self-report their reputation \cite{SORI,Ganeriwal}. Therefore, such methods are prone to the security and reliability of communication channels to report these reputations. Another vulnerability of such methods is the possibility of reporting false reputations by selfish/malicious nodes that requires an audit unit to identify such misbehaving users \cite{Soltanali,Loukas}.  Also, authors in \cite{zhao2011weighted} proposed a weighted cooperative spectrum sensing framework for infrastructure-based networks which requires a base station to receive reports and updates from secondary users. But, our model does not need any base station, so it is compatible with any infrastructureless or ad hoc networks. 
In \cite{zhang2013redisen}, a reputation-based secure method is proposed for distributed networks that involves a heavy overhead for update messages from neighbor nodes. However, in our proposed model, the reputation of the potential relay nodes is directly observed by the source node (e.g. primary node), hence it can prevent the impact of false reports. If such a first-hand reputation of a relay node was not available to a primary user, it can inquire this reputation from other primary users in the neighborhood (second-hand reputation) as further explained in Section \ref{Sec:game}.

Let us consider the problem of spectrum leasing to secondary IoT devices in the presence of a passive eavesdropper, in which these IoT devices provide cooperative relaying and cooperative jamming services to enhance the secrecy rate and quality of service of the primary user in exchange for spectrum access. Since the IoT devices often have a limited energy, the natural tendency of such non-altruistic users can often lead to selfish attacks where they violate their commitments to the primary users after they are granted the spectrum access. In this paper, we propose a reputation-based method to monitor the cooperative behavior of these IoT SUs in terms of the power they dedicate to requested services from the primary user including cooperative jamming and cooperative relaying to prevent potential selfish behavior. 


The rest of this paper is organized as follows. In Section \ref{Sec:system}, the system model is described. The formulation of the proposed reputation-based Stackelberg game is defined in Section \ref{Sec:game}. The simulation results are presented in Section \ref{Sec:results}, followed by concluding remarks in Section \ref{Sec:con}.

\section{System Model} \label{Sec:system}

\begin{figure}[t]
  \vspace{-5pt}
	\centering
	\includegraphics[width=0.9\linewidth,keepaspectratio]{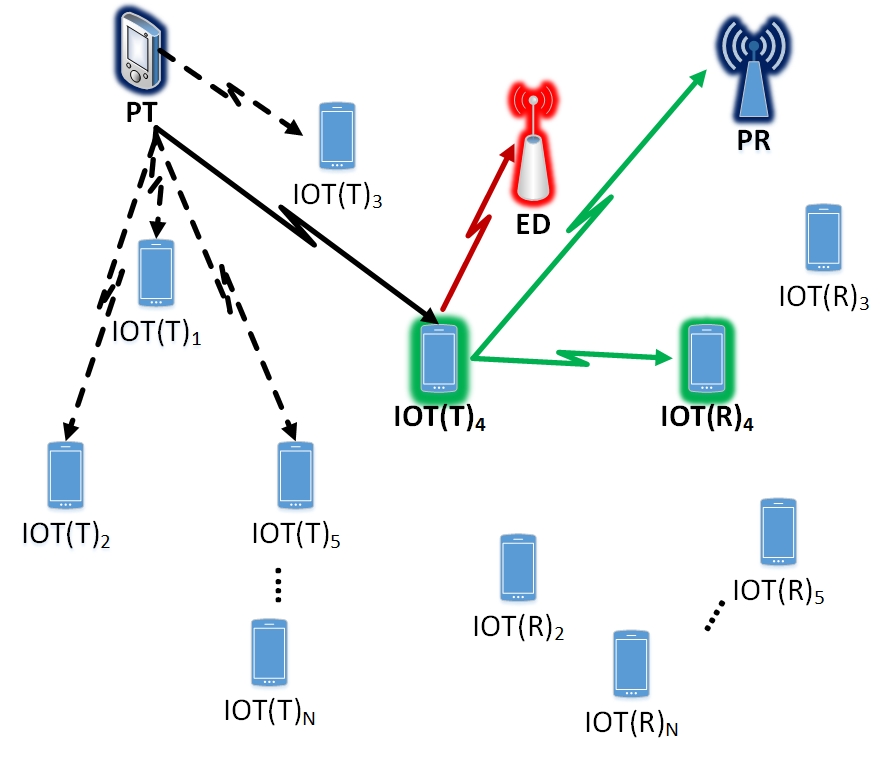}
	\caption{An example of spectrum leasing to secondary IoT devices in exchange for cooperative jamming and cooperative relaying. In this example, $ST_{4}$ is the selected trustable secondary IoT.}
	\label{fig:Fig1}
    \vspace{-15pt}
\end{figure}

In this paper, we consider a cognitive radio network consists of a single primary user (PU), and $N$ secondary IoT users (SUs) who are seeking to obtain spectrum access. There also exists a passive malicious eavesdropper, $ED$, who attempts to decode the PU's message, as depicted in Fig. \ref{fig:Fig1}. The PU's transmitter, $PT$, intends to send a secure message to its receiver, denoted by $PR$. The transmitter and receiver of SU $i$ are denoted by $ST_i$, and $SR_i$, respectively. All transmitters and receivers are assumed to have a single antenna. 

It is assumed that channels between the nodes are slow Rayleigh fading with constant coefficients over one time slot. These channel coefficients are defined as follows: i) $h_{PS_i}$ refers to channel coefficients between the PU's transmitter and $i^{th}$ SU's transmitter; ii) $h_{S_iP}$, $h_{S_i}$, and $h_{S_iE}$ denote the channel coefficients between $i^{th}$ SU transmitter and the PU's receiver, the $i^{th}$ SU receiver, and the eavesdropper, respectively. We assume that there is no direct link between $PT$ and $PR$. A common assumption of availability of Channel State Information (CSI) at the transmitters using standard channel state estimation techniques is followed here \cite{Stanojev_Secrecy,Liu_Secrecy_Game,Talabani}. The noise sources at the receivers are complex and circularly symmetric, i.e. $n \sim CN(0,\sigma^2)$. A constant power $P_P$ is assumed at the PU, while the maximum available energy for each time slot at SU $i$ is $E_i^{max}$. 

\begin{figure}[b]
	\centering
      \vspace{-5pt}
	\includegraphics[width=\linewidth,keepaspectratio]{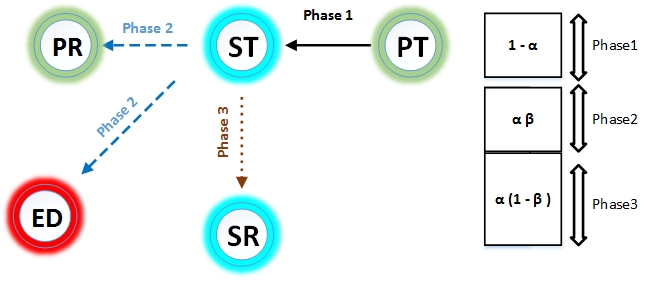}
	\caption{System model of proposed spectrum leasing scheme between the PU and the selected trustable secondary user.}
	\label{fig:Fig2}
    \vspace{-15pt}
\end{figure}

Each time slot of PU's access to the radio spectrum, $T$, is divided to three phases, as depicted in Fig. \ref{fig:Fig2}. During Phase I, $(1-\alpha)T$, the PU broadcasts its message to the available secondary users. It is assumed that the eavesdropper is out of the range of direct communication from the PU's transmitter.\footnote{If the transmission in Phase I was subject to eavesdropping, the primary user can allocate a portion of its power to sending an artificial noise to protect its transmitted signal from the eavesdropper. } If the transmitted signal at $PT$ is denoted by $S$, the received signal at SU $i^{{th}}$ transmitter during the first phase is 
\begin{equation}
X_{\text{ST}_i} = \sqrt{P_P}h_{PS_i}S + n_{ST}
\end{equation}
where  $n_{ST} \sim CN(0,\sigma^2)$ denotes the noise at $ST_i$.

During Phase II, $\alpha\beta T$, the selected trustable secondary user (as defined in Section \ref{Sec:game}) participates in cooperative relaying and jamming. If secondary user $i$ is selected by the PU, it forwards the PU's message to $PR$ with power $P_{i_C}$, and also adds artificial noise $z$ with power $P_{i_J}$ to confuse the eavesdropper. A Decode-and-Forward (DF) relaying scheme is deployed at the secondary users where the received message is fully decoded. We assume that a secure information transmission mechanism is in place where the PU's receiver can obtain a priori knowledge of the added artificial noise by the secondary user \cite{Koorapaty,Talabani,Jafarkhani}. One practical method to obtain this a priori knowledge of artificial noise is using a pseudo-random generator with finite states at the relay and legitimate primary receiver, while the state of the pseudo-random generator is sent to the $PR$ via a secure control channel. While this mechanism only involves a small amount of overhead, it results in enhancing the secrecy of the PU's transmission since the artificial noise creates interference at the eavesdropper but it can be fully removed at the $PR$. Hence, the received signal at $PR$ can be written as
\begin{equation}
X_{PR} = \sqrt{P_{i_C}}h_{S_iP}\hat{S} + n_{PR}
\end{equation}
where $\hat{S}$ denotes the re-encoded signal transmitted by $ST_i$ after DF, and $n_{PR} \sim CN(0,\sigma^2)$ is noise at $PR$. The received message at the eavesdropper during Phase II is: 
\begin{equation}
X_{ED} = \sqrt{P_{i_C}}h_{S_iE}\hat{S} + \sqrt{P_{{i_J}}}h_{S_iE}z + n_{ED}
\end{equation}
where $z \sim CN(0,1)$ is the artificial noise added by $ST_i$.

During Phase III, the selected secondary user transmits its own message to its corresponding destination with power $P_{i_S}$. If the transmitted signal by $ST_i$ is denoted $\hat{S_i}$, the received signal at $SR_i$, $X_{S_iR}$ can be written as 
\begin{equation}
\nonumber
X_{S_iR} = \sqrt{P_{i_S}}h_{S_i}\hat{S_i} + n_{S_iR}
\end{equation}
$n_{S_iR} \sim CN(0,\sigma^2)$ is the noise at $SR_i$.

Based on this model, if the codewords at the PU are Gaussian inputs, the achievable secrecy rate of the primary user, $R_{P_{\text{SEC}}}$, defined as the communication rate at which no information is revealed to the eavesdropper can be written as
\begin{align}
R_{P_{\text{SEC}}}=[R_{P}-R_{E}]^+
\end{align}
where $R_{P}$ is the accumulated rate at the PU's receiver and $R_E$ is the accumulated rate at the eavesdropper (information leakage), and $[x]^+$ denotes $max(x,0)$. For simplicity of notation, we remove $[.]^+$ sign from now on. The rate at the PU's receiver is defined as $R_P=\min(R_{PS_i},R_{S_iP})$ noting the DF relaying scheme at the selected secondary user $i$ \cite{Laneman}, hence the secrecy rate at $PR$ can be calculated as:
\begin{align} \label{RSEC}
&R_{P_{\text{SEC}}}=\min \big ((1-\alpha)T \log_2(1+\frac{P_P|h_{PS_i}|^2}{\sigma^2}),\\
\nonumber
&\alpha\beta T \log_2 (1+\frac{P_{{i_C}} |h_{S_iP}|^2 }{\sigma^2})\big)\\
\nonumber
&-\alpha \beta T \log_2(1+\frac{P_{{i_C}}|h_{S_iE}|^2}{\sigma^2+ P_{{i_J}}|h_{S_iE}|^2})) 
\end{align}
It is worth mentioning that to achieve nonzero secrecy rate for the primary user, the  conditions of $|h_{PS_i}|^2 >|h_{S_iE}|^2, |h_{S_iP}|^2 > |h_{S_iE}|^2$ should be satisfied for the CSIs, where there is no direct link between the primary users and eavesdropper \cite{Zhang_Wyner}.


\section{Proposed Reputation-based Stackelberg Game Model} \label{Sec:game}
Here, we define a Stackelberg game model to describe the interactions between the primary and secondary users in the proposed spectrum leasing scenario. The objective of the PU is to enhance its secrecy rate through utilizing the reliable secondary nodes while the secondary IoT users aim to obtaining the chance of spectrum leasing from the PU and maximize their transmission rate considering the power cost of cooperation with the PU. Such cost refers to the consumed power for forwarding the PU's message and adding a jamming signal to it. Despite the previously reported works, where the SUs were forced to utilize an equal amount of power for their own transmission and service for the PU \cite{Talabani,Stanojev_Secrecy,Liu_Secrecy_Game}, we account for the selfish nature of the SU's behavior to allocate a small portion of their power $(P_{i_J}+P_{i_C})$ to the PU's requested services, in order to maximize their own transmission rate by having a larger remained power to themselves, $(P_{i_S})$. It is worth mentioning that both services requested by the PU (jamming and relaying) occur during the second phase ($\alpha \beta T$), hence the selected SU does not benefit from changing the ratio of the power spent on relaying or jamming since its overall cost of service to the PU remains as $(P_{i_J}+P_{i_C}) \alpha \beta T$. Therefore, it is reasonable to assume that the ratio of power for jamming and relaying denoted by $\rho=\frac{P_{i_J}}{P_{i_C}}$ is requested by the PU and the selected SU has no incentive to change this ratio.

In the proposed Stackelberg game, the PU as the spectrum owner is defined as the game leader to declare its strategies first, and the SUs are the game followers. The strategy set of the PU is to: i) select the most reliable secondary user; and ii) determine the optimal time allocation among the three phases, i.e parameters $\alpha$ and $\beta$.
A reputation value is assigned to each $SU$ that keeps track of their performance in terms of power allocation between their own transmission and requested services by the $PU$. The reputation value of each SU is monitored and maintained by the PU, and increases when the $SU$ allocates enough power for cooperation and jamming. Let us define $R^k(n)=(r_1(n),r_2(n), ..., r_N(n))$ as the reputation vector stored at the $k^{th}$ PU at time $n$, where $r_i^k(n) \in (0,1]$ denotes the reputation of SU $i$ observed by $k^{th}$ PU at time $n$ for $i \in \{1,2,...,N\}$.
Here, we define \emph{first-hand reputation} and \emph{second-hand reputation} based on the existence of prior interactions between a primary and secondary pair. If PU $k$ has a history of direct interaction with candidate SU $i$, the first-hand reputation at time $n$, $r_i^k(n)$ is defined as 
\begin{align} \label{reputation}
r_{i}^{k}(n) = \min(r_{i}^{k}(n - 1) + \eta_{3}\ln(\epsilon_i^n),1)
\end{align}
where $\eta_{3}$ is the pre-defined factor and $\epsilon_i^n$ is a function of power for individual transmission of user $i$, the power for requested services by the $PU$, and the channel conditions at time $n$, i.e.  ($\epsilon_i^n=\frac{P_{i_J}|h_{S_iE}|^2+P_{i_C}|h_{S_iP}|^2}{P_{i_S}|h_{S_i}|^2}$). The reason behind considering the CSIs in this definition is to assign a change of reputation proportional to actual channel condition that a secondary user is experiencing. With this definition, the reputation of misbehaving secondary users declines rapidly due to the characteristic of $\ln$ function while it would take a fairly long time to restore the reputation in case of misbehaving. 
This mechanism can potentially prevent the secondary users from oscillating between good behavior and misbehavior due to the long period of time required to restore their reputation. 

If a user recently moves to a proximity of a primary user or there does not exist a prior record of interactions between a primary and secondary pair, then the primary user can inquire the SU's reputation from its neighbor. In these cases, the second-hand reputation is defined as
\begin{align}
r_i^k(n)=\frac{\Sigma_{j\in \mathcal{P}_i^k} r_i^j (n)}{|\mathcal{P}_i^k|}
\end{align}
where $\mathcal{P}_i^k$ is the set of primary users in proximity of $k^{th}$ PU. If a secondary user is new in the network and there is no prior information about its reputation, then the reputation of this unit will be set to the lower bound of reputation range. 

\begin{figure}
  \vspace{-5pt}
	\centering
	\includegraphics[width=\linewidth,keepaspectratio]{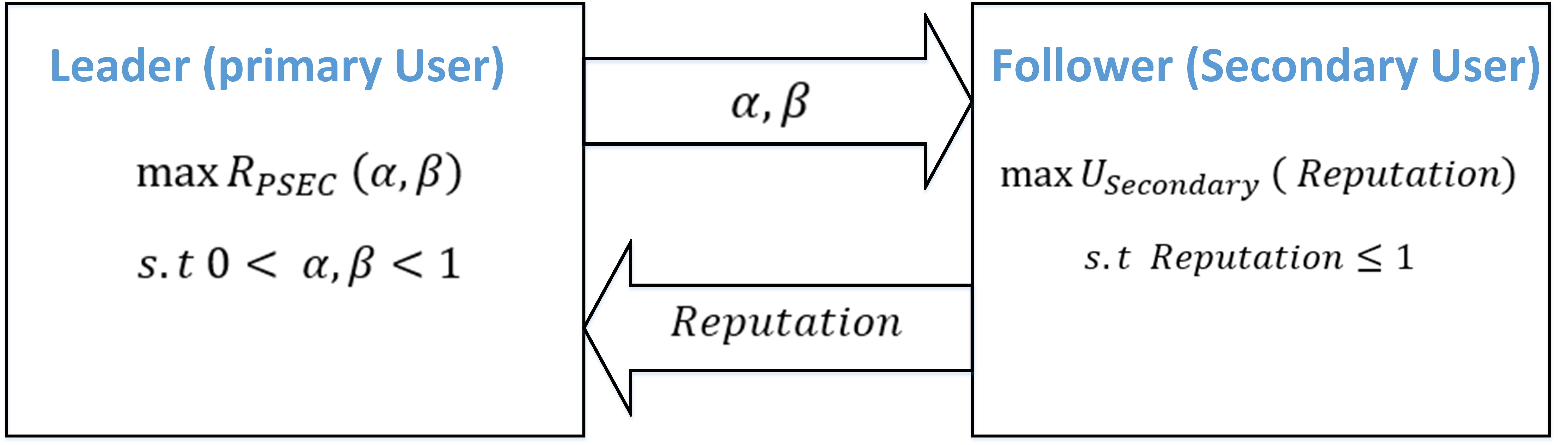}
	\caption{Stackelberg Game Model.}
	\label{fig:Fig3}
       \vspace{-15pt}
\end{figure}

The utility function of secondary user $i$ is defined as:
\begin{align} \label{eq:Us_0}
&U_{S_i} (P_{i_S},P_{i_C},P_{i_J}) = \alpha (1 - \beta) T log_2(1+{SNR}_{i}) \\
\nonumber
&- \eta_{1}\alpha(1-\beta) T P_{i_S} -\eta_{2}\alpha\beta T (P_{i_C} +P_{i_J}) \\
\nonumber
&+ \ln(P_{i_J}|h_{S_iE}|^2+P_{i_C}|h_{S_iP}|^2 - P_{i_S}|h_{S_i}|^2 + 1)
\end{align}
where $SNR_i$ denotes the signal-to-noise ratio at $i^{th}$ SU receiver. The last term in (\ref{eq:Us_0}) is proportional to the change of reputation. Therefore, decision making at secondary user $i$ involves a trade-off to increase its own transmission power $P_{i_S}$ to obtain a higher transmission rate; while it needs to dedicate enough power to cooperation, $P_{ic}$ and jamming, $P_{ij}$ to maintain a good reputation in order to be chosen for the next rounds.
The condition related to limited energy at SU $i$ can be written:
\begin{align} \label{U_s:condition}
&\alpha\beta T P_{i_C} + \alpha\beta T P_{i_J} + \alpha(1-\beta)TP_{i_S} \leq \alpha T P_{i_\text{max}}
\end{align}
where $P_{i_\text{max}}$ denotes the maximum power available at SU $i$. Since these secondary IoT users are self-interested, they exhaust their available energy that results in $\beta P_{i_C} + \beta P_{i_J} + (1-\beta)P_{i_S} =  P_{i_{\text{max}}}$. Noting this equality and the requested ratio of power for jamming and cooperation by the PU, $\rho=\frac{P_{i_J}}{P_{i_C}}$, the utility of SU $i$ can be written as a function of one of the three power strategies of this user, (e.g. $P_{i_S}$) as
\begin{align} \label{eq:Us}
&U_{S_i} (P_{i_S}) = \alpha (1 - \beta) T log_2(1+\frac{P_{i_S}|h_{s_i}|^2}{\sigma^2}) \\
\nonumber
&- \eta_{1}\alpha(1-\beta) T P_{i_S} -\eta_{2}\alpha\beta T \frac{P_{i_\text{max}}-P_{i_S}(1-\beta)}{\beta(1+\rho)} \\
\nonumber
&- \eta_{2}\alpha\beta T \rho \frac{P_{i_\text{max}}-P_{i_S}(1-\beta)}{\beta(1+\rho)}\\
\nonumber
&+ \ln(P_{i_J}|h_{S_iE}|^2+P_{i_C}|h_{S_iP}|^2 - P_{i_S}|h_{s_i}|^2 + 1)
\end{align}
where $\eta_{1}$ and $\eta_{2}$ are predefined normalizing coefficients for energy to form it similar to transmission rate. The hierarchical interactions between the Stakelberg game players are summarized in Fig \ref{fig:Fig3}. 
The PU's optimum parameters ($\alpha^*,\beta^*$, k), and the corresponding power choice of the SU ($P_{i_S}$), are cooperatively considered as the Stackelberg solution that can be obtained through a backward induction algorithm. In each iteration of the game, the selected SU observed the selected strategy of the PU (i.e. $\alpha$, $\beta$), and responded to this by selecting its optimum power allocation to maximize its utility.

\textbf{Lemma 1}: The utility of secondary user, (\ref{eq:Us}) is concave in term of $P_{i_S}$.\\
$Proof$: To prove the concavity of secondary utility, the second derivative of (\ref{eq:Us}) with respect to $P_{i_S}$ is driven as:\\
\begin{align} \label{concave:U_S}
&\frac{\partial^{2}U_s(P_{i_S})}{\partial^2 P_{i_S}} = \frac{-\alpha(1-\beta)T}{\ln(2)}\frac{(\frac{|h_{s_i}|^2}{\sigma^2})^2}{(1+\frac{P_{i_S}|h_{s_i}|^2}{\sigma^2})^2}\\\nonumber
& +  \frac{-A^2}{(P_{i_C}|h_{S_iP}| + P_{i_J}|h_{S_iE}| - P_{i_S}|h_{s_i}| + 1)^2} < 0
\end{align}
where $A^2$ is equal to $ (-(\frac{(1-\beta)}{\beta(1+\rho)})|h_{S_iP}| - (\rho\frac{(1-\beta)}{\beta(1+\rho)})|h_{S_iE}| -|h_{s_i}|)^2 $. As seen in (\ref{concave:U_S}), $U_{S_i}$ is strictly concave in terms of $P_{i_S}$ for given values of $\alpha$, and $\beta$. 

\textbf{Theorem 1}: The Stackelberg solution of the proposed game is the unique Nash equilibrium.\\
\textbf{Proof}: During the backward induction process to obtain the Stackelberg solution, the strategy of the PU is selected by maximizing (\ref{RSEC}) as $
\alpha^*, \beta^* = \text{argmax} R_{PSEC}(\alpha,\beta)$. Then the selected SU observes this strategy and selects its optimum power allocation set. According to \emph{lemma 1}, the utility of the selected SU is strictly concave in terms of $P_{i_S}$ for any given value of the PU's strategy set, therefore the response of the selected SU is unique. It is obvious that the proposed single-leader single-follower Stackelberg game converges to the obtained solution as the follower is rational and the PU can fully predict the optimal strategy of the selected SU. Moreover, since the strategy of each user is obtained through finding the best response to the strategy of the other one, the obtained Stackelberg solution is a Nash equilibrium. 




\section{Simulation Results} \label{Sec:results}
In this section, we present the simulation results to evaluate the performance of the proposed reputation-based game model in enhancing the secrecy rate of the primary user through two scenarios. The channels between nodes $i$ and $j$ are obtained from  $h_{i,j} \sim CN(0,d^{-2}_{i,j})$ where $d_{i,j}$ is the distance between nodes $i$ and $j$. 
The duration of one time slot, $T$, is assumed to be equal to 1. The values of $\eta_{1}$ and $\eta_{2}$ are set to $0.004$ and $0.0005$, respectively. To have a more strict penalization for selfish behavior of the SUs, we assume that $\eta_{1}$ is higher than $\eta_{2}$. The variance of noise ($\sigma^2$) is equal to $1$mw, $\rho$ is $0.7$, and the primary transmitter uses $3$mw power for sending its packet to the secondary user, unless explicitly stated otherwise.

\subsection*{Scenario 1: Single Secondary User}
\begin{figure}
	\centering
	\includegraphics[width=\linewidth,keepaspectratio]{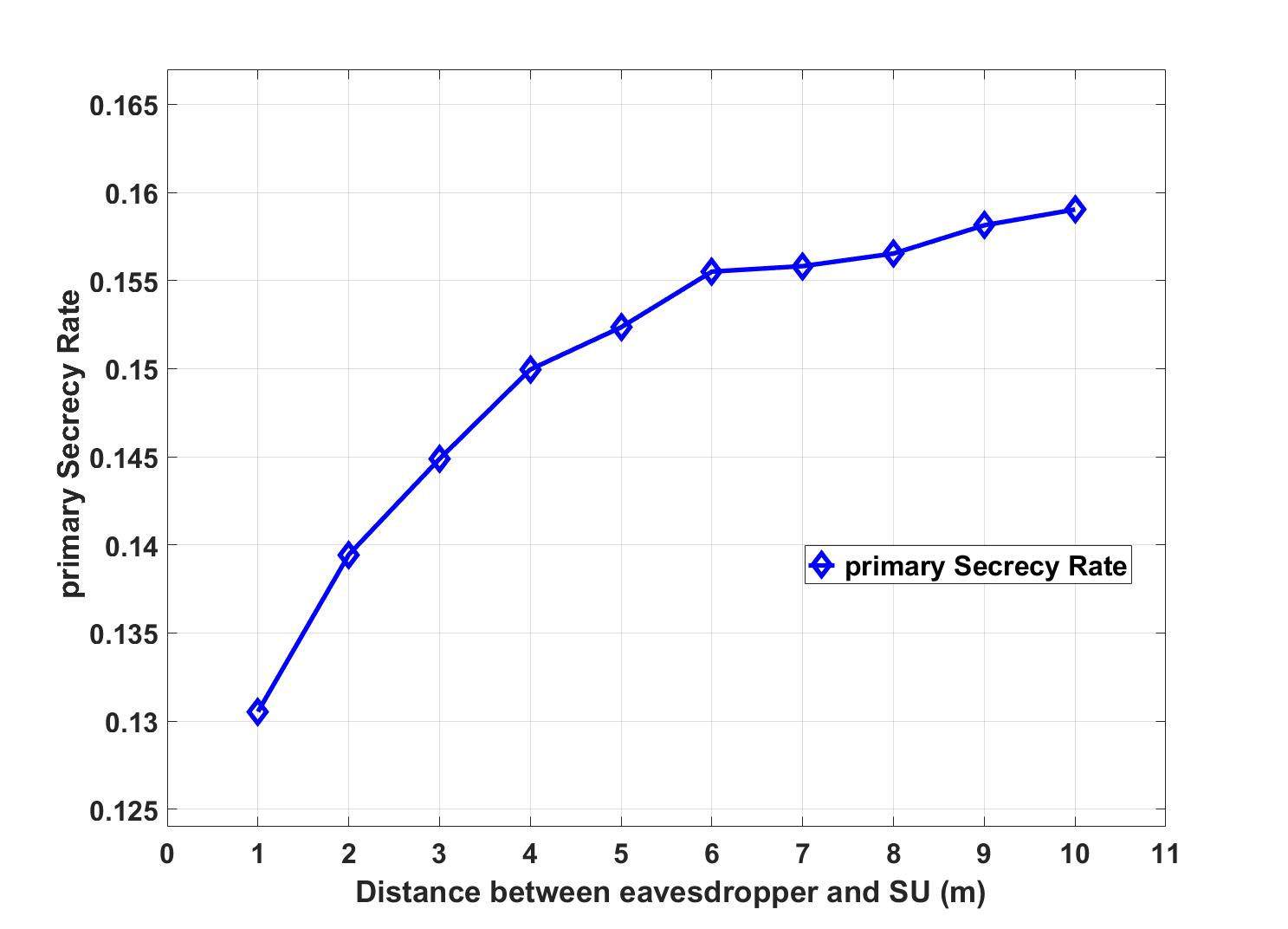}
	\caption{Primary Secrecy Rate versus distance of eavesdropper and ST}
	\label{fig:Fig4}
    \vspace{-15pt}
\end{figure}

\begin{figure}
  \vspace{-5pt}
	\centering
	\includegraphics[width=\linewidth,keepaspectratio]{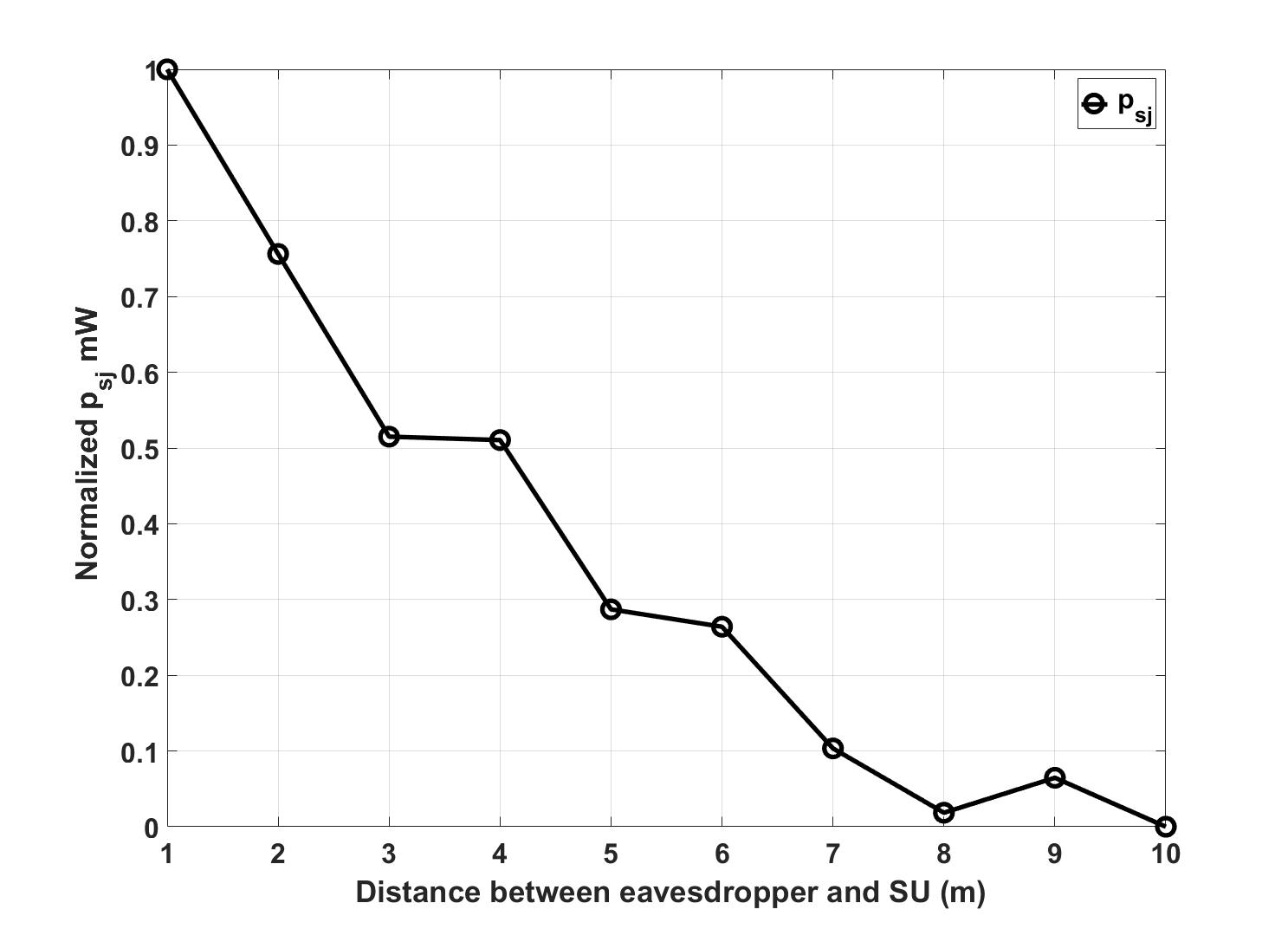}
	\caption{Jamming power of secondary versus distance of eavesdropper}
	\label{fig:Fig5}
    \vspace{-15pt}
\end{figure}

In this scenario, there exists only one SU in the neighborhood of the PU. 
Fig. \ref{fig:Fig4} illustrates the primary secrecy rate for different distances of eavesdropper with the relay. As shown in this figure, increasing the distance between the relay and eavesdropper results in increasing the PU's secrecy rate. It also results in decreasing the jamming power of the secondary transmitter as the communication will be less vulnerable to the presence of the eavesdropper, as depicted in Fig \ref{fig:Fig5}.
  \vspace{-5pt}

\subsection*{Scenario 2: Multiple Selfish and Reliable Secondary Users}
\begin{figure}[b]
  \vspace{-10pt}
	\centering
	\includegraphics[width=\linewidth,keepaspectratio]{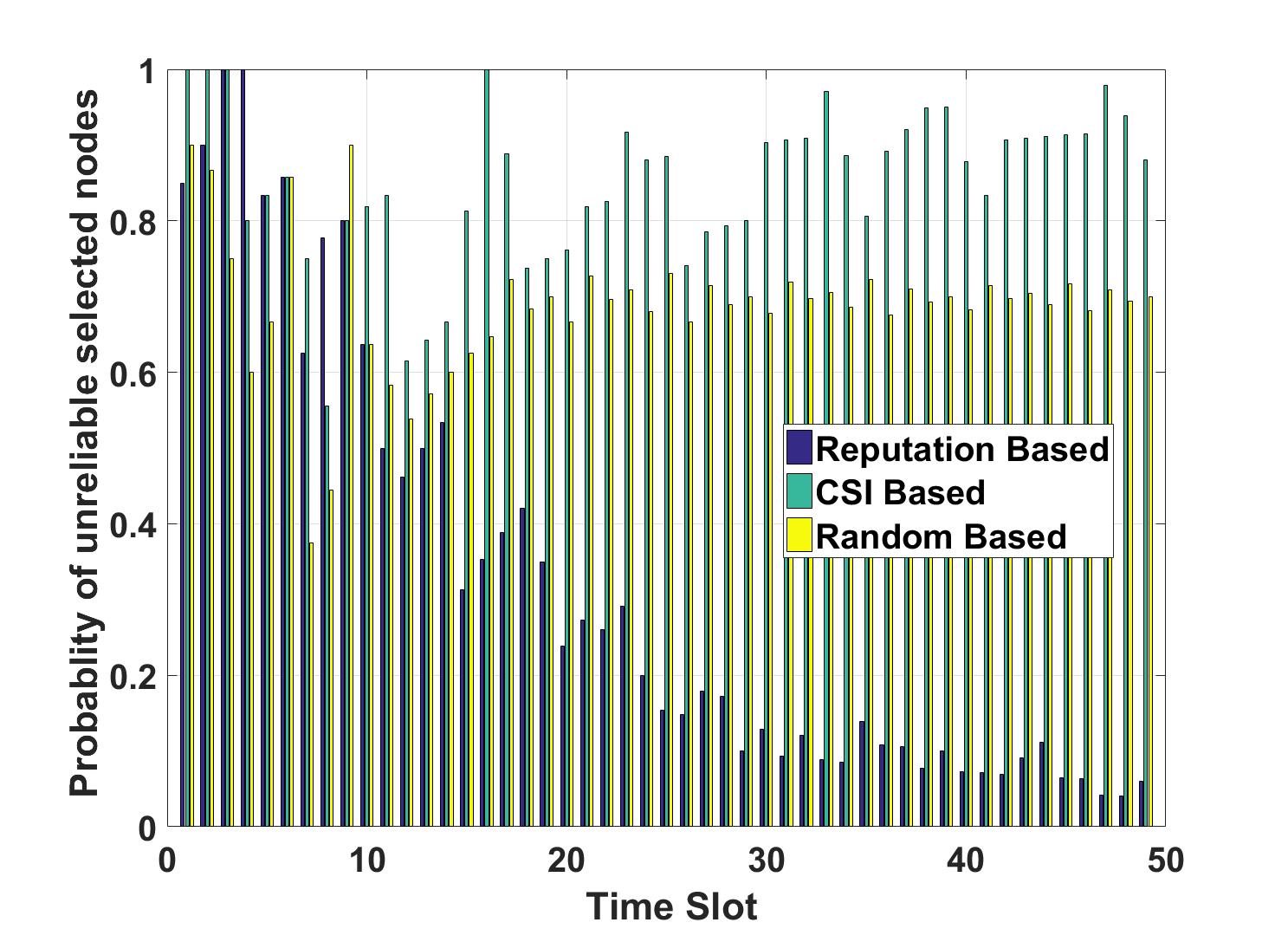}
	\caption{Probability of selecting unreliable nodes over time}
      \vspace{-5pt}
	\label{fig:Fig6}
    \vspace{-15pt}
\end{figure}

In this scenario, we consider $10$ mobile SUs, in which $70\%$ of them are selfish in the sense that they may assign all their energy to their own transmission if granted with the spectrum, and the rest of the SUs are fully reliable. It is assumed that the location of SUs changes in every time slot to evaluate the performance of the proposed method in a more practical scenario. The SUs are  uniformly distributed in the neighborhood around the PU. The performance of our proposed reputation-based method is compared with random relay selection, and best CSI relay selection methods. In all three methods, Stackelberg solution is found as the solution of the mechanism.  In order to demonstrate the effectiveness of our method related to existing techniques, we considered an extreme case where the selfish secondary nodes are closer to the PU compared to the reliable ones. Also, it is assumed that selfish SUs show selfish behavior in $20\%$ of time slots, hence it is less likely for the PU to identify all selfish nodes using alternative techniques. Fig. \ref{fig:Fig6} demonstrates the probability of selecting unreliable nodes for different selection methods. As seen in this figure, the best CSI-selection method, which is widely used in many relay selection applications, has the worst performance in this scenario because it chooses the nearest SUs. For random selection method, the probability of selecting the unreliable SUs is close to $70\%$ and does not change over time, as expected. However, our proposed reputation-based model is capable of filtering the selfish users over the course of time.

\begin{figure}
  \vspace{-5pt}
	\centering
	\includegraphics[width=\linewidth,keepaspectratio]{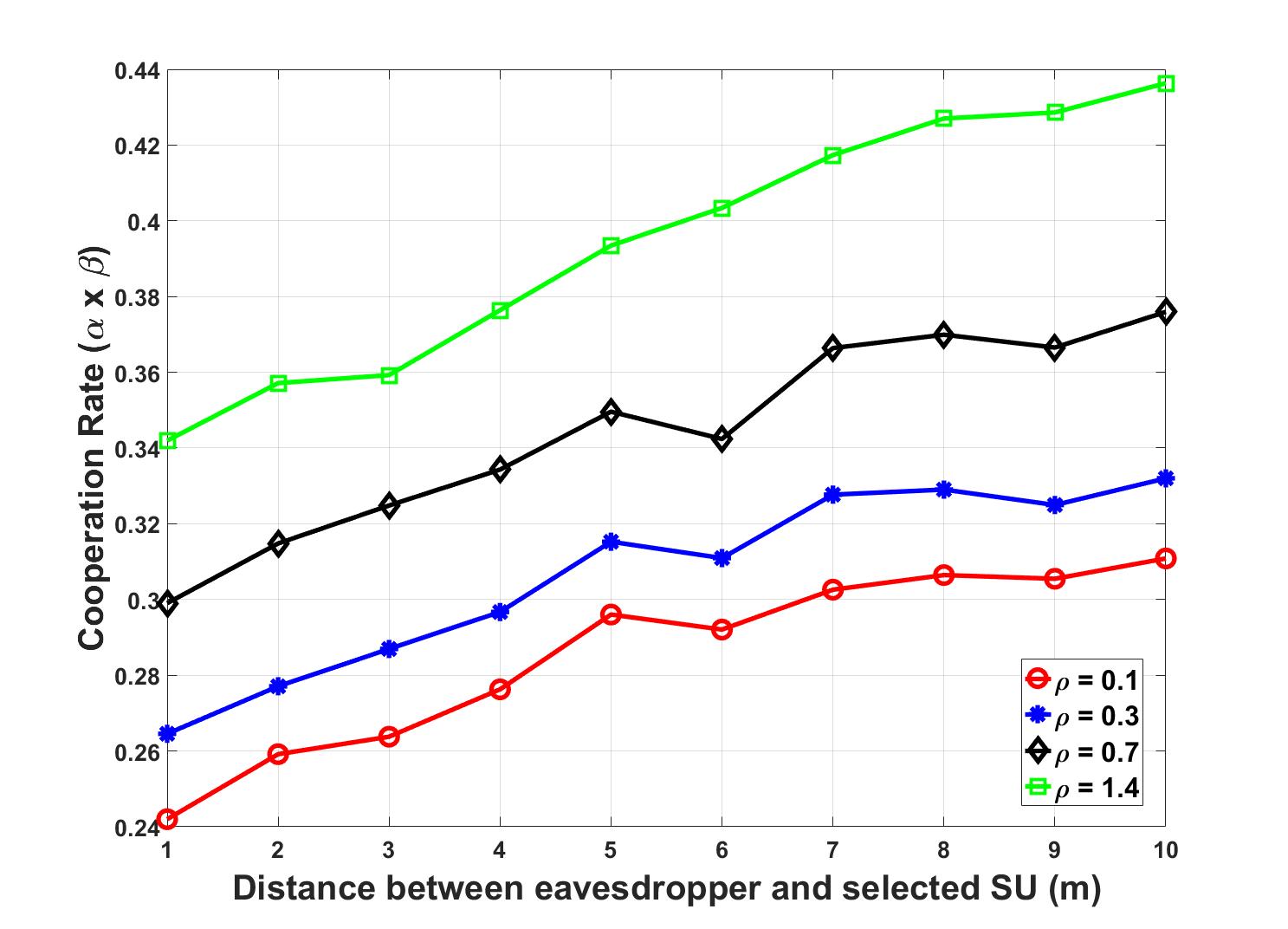}
	\caption{cooperation phase ($\alpha\beta$) versus the distance between eavesdropper and the selected SU for different values of $\rho$}
	\label{fig:Fig12}
      \vspace{-15pt}
\end{figure}

Fig. \ref{fig:Fig12} shows the allocated time to cooperative services, $\alpha \beta$, determined by the PU versus the distance between the eavesdropper and selected SU. The PU is encouraged to assign more time to cooperative services when the cooperative link is less vulnerable to eavesdropping in order to enhance its quality of communication through relaying services provided by the SU.

\section{Conclusion} \label{Sec:con}
In this paper, we proposed a practical solution for spectrum leasing to unlicensed IoT devices, in which the spectrum incumbents are motivated to lease their spectrum through receiving cooperative service from the IoT devices as well as enhancing their secrecy rate via a cooperative jamming mechanism. The key contribution of this model is to develop a reputation-based mechanism that enables the spectrum owners to monitor the behavior of potentially selfish IoT devices in terms of their willingness and efforts to participate in cooperative services. In the majority of existing spectrum sharing solutions, the secondary users are assumed to be fully reliable, hence the spectrum owners can often suffer from the selfish attacks performed with self-interested unlicensed users. However, our proposed spectrum leasing method gives the spectrum owners the possibility of only interacting with the reliable IoT devices and prevents the chance of degradation in quality of service due to selfish attacks, as confirmed by the simulation results.



\section{ACKNOWLEDGMENT OF SUPPORT AND DISCLAIMER}
The authors acknowledge the U.S. Government's support in  the publication of this paper. This material is based upon work
funded by AFRL, under AFRL Grant No. ICA2017-SVFRP-011. Any opinions, findings and conclusions or recommendations
expressed in this material are those of the author(s) and do not necessarily reflect the views of the US government or AFRL.

  \vspace{-5pt}
\bibliographystyle{IEEEtran}
\bibliography{References_Sec}

\end{document}